\documentclass[12pt,twocolumn]{iopart}
\usepackage[dvips]{graphicx}
\usepackage{array}
\usepackage{bm}
\usepackage{fancyhdr}
\usepackage{makeidx}
\usepackage{epsfig}
\usepackage{amsfonts}
\usepackage{amssymb}
\usepackage{color}
\begin{document}

\title[Trend of the magnetic anisotropy for Mn dopants]{Trend of the magnetic anisotropy for individual Mn dopants near the (110) GaAs surface} 
\author{M R Mahani, A Pertsova and C M Canali} 
\address{Department of Physics and Electrical Engineering,
Linn{\ae}us University, Norra V{\"{a}}gen 49, 391 82, Kalmar, Sweden}
\ead{carlo.canali@lnu.se}

\begin{abstract} 
Using a microscopic 
finite-cluster tight-binding model, 
we investigate the trend of the magnetic anisotropy energy 
as a function of the cluster size for an individual Mn impurity  
positioned in the vicinity of the (110) GaAs surface, 
We present results of calculations 
for large cluster sizes, containing approximately  $10^4$ atoms, which have not been 
investigated so far. Our calculations demonstrate that the anisotropy energy of 
a Mn dopant in bulk GaAs found to be non-zero in previous tight-binding calculations,
is purely a finite size effect and it vanishes as the inverse cluster size.
In contrast to this, we find that the splitting of the three 
in-gap Mn acceptor energy levels
converges to a finite value in the limit of infinite cluster size. For a Mn in bulk GaAs
this feature is related to the nature of the mean-field treatment of the coupling between the 
impurity and its nearest neighbors atoms. 
Moreover, we calculate the trend of the anisotropy energy in the sublayers, as the Mn dopant 
is moved away from the surface towards
the center of the cluster. Here the use of large cluster sizes allows us to position 
the impurity in deeper sublayers below the surface, compared to previous calculations. 
In particular, we show that the anisotropy energy increases up to the fifth sublayer 
and then decreases as the impurity is 
moved further away from the surface, approaching its bulk value.  
The present study provides important insight for experimental control and manipulation 
of the electronic and magnetic properties of individual Mn dopants at the semiconductor surface by means 
of advanced scanning tunneling microscopy techniques.
\end{abstract}
 
\maketitle

\section{\label{sec:level1}Introduction\protect}
\begin{figure}[htp]
\centering
\includegraphics[width=0.9\textwidth]{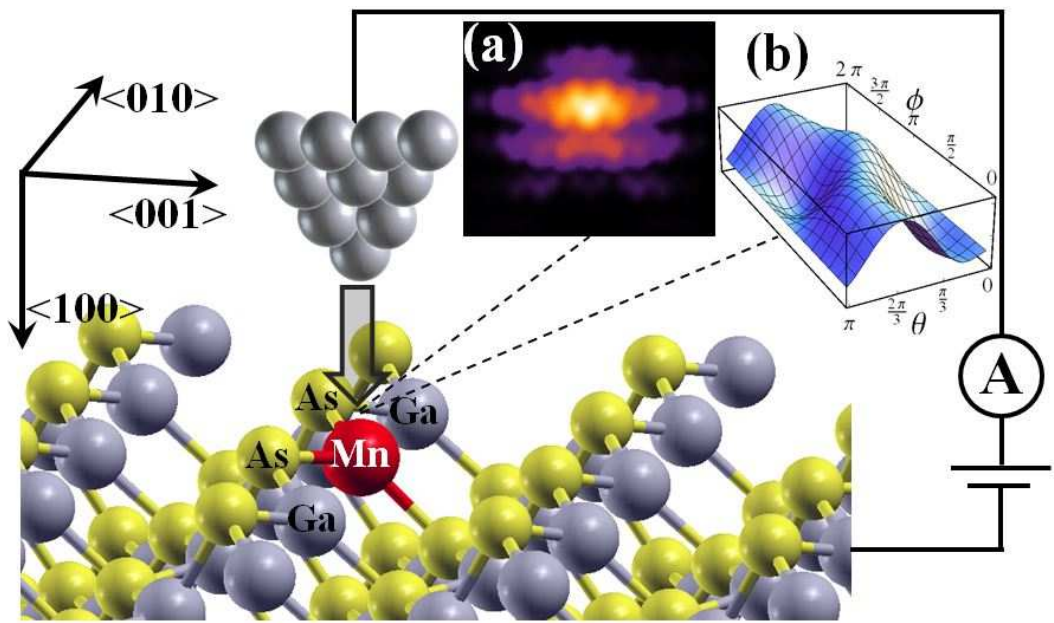}
\caption{Color online -- Schematic of an STM experiment on single Mn impurity positioned on the (110) surface of GaAs.  
The insets show two examples of electronic and magnetic properties of the Mn dopant, which can be calculated using the TB model 
employed in this paper: (a) the calculated LDOS for the Mn acceptor state and (b) the calculated magnetic anisotropy landscape 
for Mn in the surface layer.}
\label{fig:intro}
\end{figure}

The past decade has witnessed a surge of interest in understanding and actively controlling 
electronic, optical and magnetic properties of solitary dopants in semiconductors.  
A corresponding sub-division of semiconductor electronics, known as \textit{solotronics} (solitary dopant optoelectronics), 
has emerged in recent years, with the focus on building novel devices that would make use of specific properties of 
individual dopants, as well as on advancing our fundamental understanding of these atomic-scale systems~\cite{pm_nam11}. 
The experimental progress in this field has been largely driven by 
remarkable advances in using scanning tunneling microscopy (STM) to custom engineer, manipulate and characterize 
single impurities on surfaces with atomic precision~\cite{Hirjib06, yazdani_nat06}. 
In a number of key experiments, STM based techniques were used to study the electronic structure and 
the magnetic interactions of 
substitutional transition-metal (TM) dopants at semiconductor surfaces~
\cite{yakunin_prl04, shinada_nat05, yazdani_nat06, 
wiesendanger_MnInAs, koenraad_prb08, gupta_science_2010, garleff_prb_2010,
Fuec_nnt12, pla_nat12}. 
%The enhanced stability of the STM at low temperature
%allows layer by layer identification of Mn atoms embedded in the 
%first few atomic layers of the GaAs crystal~\cite{koenraad_prb08}.
On the theoretical side, both first-principles calculations~\cite{zhao_apl04,sarma_prl04, 
PhysRevB.70.085411, PhysRevB.75.195335, ebert_prb09, fi_cmc_prb_2012} 
and microscopic tight-binding (TB) 
models~\cite{tangflatte_prl04, tangflatte_prb05, timmacd_prb05,Jancu_PRL_08, scm_MnGaAs_paper1_prb09, 
scm_MnGaAs_paper2_prb2010, scm_MnGaAs_paper3_prl011, PhysRevLett.105.227202,mc_MF_2013} have 
played an essential role in elucidating experimental findings and predicting new properties. 
 Computationally efficient and physically motivated  
TB models have been particularly successful in describing electronic and magnetic properties of some TM 
impurities, such as Mn dopants with their associated acceptor states~\cite{tangflatte_prl04, tangflatte_prb05, 
Jancu_PRL_08, scm_MnGaAs_paper1_prb09,
scm_MnGaAs_paper2_prb2010, scm_MnGaAs_paper3_prl011, mc_MF_2013}  
and, more recently, Fe dopants~\cite{rm_d_level} on the (110) GaAs surface.\\
Due to their computational feasibility, microscopic TB models are especially well suited to study 
single impurities as they allow the use of large supercells, with sizes exceeding  
those accessible by first-principles approaches by several orders of magnitude. 
Such models allow the calculation of measurable physical quantities, which can be 
directly probed  
in experiments (see Figure~\ref{fig:intro}). In particular,  {\it finite-cluster} TB calculations 
provide a detailed description of the in-gap electronic structure in the presence of the dopant close to the
surface, which can be directly 
related to resonances in conductance spectra measured by STM~\cite{scm_MnGaAs_paper1_prb09, rm_d_level}. 
Although a more elaborate treatment is required for 
simulations of STM topographic images, in the first approximation the tunneling current is 
proportional to the local density of states (LDOS) at the surface~\cite{STM_book}. Therefore,   
typically there is a strong correlation between the calculated LDOS 
maps for dopants positioned on the surface or in subsurface layers and the  
corresponding STM topographies~\cite{tangflatte_prl04, tangflatte_prb05, 
Jancu_PRL_08, scm_MnGaAs_paper1_prb09,
scm_MnGaAs_paper2_prb2010, scm_MnGaAs_paper3_prl011, mc_MF_2013}. Moreover, calculations 
of magnetic anisotropy energy of TM dopants, which are accessible with current TB approaches,   
provide an important input for interpreting and predicting the results of on-going experiments, aimed 
at manipulating the magnetic moment of the dopant, e.g. by means of an external magnetic field. 
Recently, TB calculations of the magnetic anisotropy landscape, combined with analysis of the shape and 
the spatial extent of the acceptor 
wavefunction, have been used to explain experimental results on magnetic-field  
manipulation of a single Mn acceptor near the (110) GaAs surface~\cite{mc_MF_2013}. 
A similar strategy has been used to predict the effect of the magnetic field	on the magnetic moment of Fe in GaAs 
and its dependence on the valence state of the dopant~\cite{rm_d_level}.
Here we report on recent advances in TB modeling of 
 single substitutional Mn impurities, positioned near the (110) GaAs surface. 
We use a fully microscopic tight-binding model, hereafter referred to as a \textit{quantum-spin model}, 
which includes explicitly $s$-, $p$- and $d$-orbitals of the impurity atom~\cite{rm_d_level}. 
This is in contrast to the \textit{classical-spin model} used in earlier work~\cite{tangflatte_prl04, scm_MnGaAs_paper1_prb09}, where     
the Mn impurity spin is introduced as an effective classical vector, exchange-coupled to the quantum spins 
of the nearest-neighbor As atoms. We find an overall agreement with the results of previous work, in particular with 
the classical-spin model of reference~\cite{scm_MnGaAs_paper1_prb09}. Among the key features that have been already 
reported in \cite{scm_MnGaAs_paper1_prb09} and 
that are well reproduced with our present model, are (\textit{i}) the strongly localized and anisotropic 
character of the mid-gap Mn acceptor state and (\textit{ii}) the dependence of the
acceptor binding energy and magnetic anisotropy energy
on the Mn position with respect to the surface. These features have been also observed 
experimentally in \cite{yazdani_nat06, gupta_science_2010}
and in \cite{garleff_prb_2010}, respectively. 

In the present paper we clarify and resolve some
outstanding theoretical and computational issues, which have not been addressed in previous work. Importantly, 
we present calculations of the in-gap level structure and magnetic anisotropy energy for both Mn 
in the bulk and on the surface for increasing cluster size. We show  
that the fictitious anisotropy energy for Mn in the bulk, found previously~\cite{scm_MnGaAs_paper1_prb09},  
is a finite-size effect caused by the limited size of the supercell used in earlier calculations. 
Here we calculate explicitly the anisotropy energy for Mn in bulk GaAs using large clusters counting 
up to $3\cdot 10^{4}$ atom. We find that the anisotropy energy tends to zero with increasing the cluster size. 
Also, by employing lager clusters for surface calculations we show that  
the surface anisotropy energy indeed mimics its bulk counterpart when Mn is positioned deep below the surface. 	

Another feature that persisted in earlier calculations for Mn in the bulk is the emergence of 
three non-degenerate levels in 
GaAs gap (one of the levels is unoccupied and is therefore interpreted as an acceptor). 
 It is known that the three levels appearing in the gap should be degenerate in the perfectly
 tetragonal environment of an impurity in bulk GaAs, even in the presence of the spin-orbit coupling~\cite{JPCM_206005}. 
 Here we show that the lifting of the 
 degeneracy in actual TB calculations is not a finite-size effect. Instead, it is related to the breaking 
of the rotational symmetry in mean-field-like treatments of the kinetic-exchange coupling between
the TM impurity $d$-levels and the $p$-levels of the nearest neighbor As atoms.
%In reference~\cite{JPCM_206005}, where a Mn is placed in the perfect tetrahedral symmetry, 
%the three levels are degenerate even in the presence of spin-orbit coupling.\\
%
Finally, we present a comprehensive study of the magnetic anisotropy energy 
of a single Mn acceptor as a function of its position in the subsurface layers. The finite-size 
effects, stemming from the limited size of the supercell in this type of 
calculations, have been identified and, to a great extent, controlled 
by systematically increasing the cluster size.  Such detailed knowledge of the  
magnetic anisotropy energy, together with the calculated LDOS of the impurity-induced states in the gap, 
are crucial for a quantitative comparison with STM experiments, especially in the presence of 
external electric and magnetic fields~\cite{mc_MF_2013,rm_d_level}.\\ 
%I would like to add a small paragraph about an STM paper here\\
%
%
The rest of the paper is organized as follows. In the next section %~\ref{theo_model}  
we describe the details of our microscopic TB approach and discuss some 
computational issues related to the use of large supercells. 
Section~\ref{results} contains the results of the calculations, namely   
 the electronic energy spectrum  and the magnetic anisotropy of  
the Mn acceptor on the (110) GaAs surface and subsurfaces for different 
 cluster size. We also provide a 
quantitative comparison with the results of the classical-spin model,   
reported previously~\cite{scm_MnGaAs_paper1_prb09},  
as well as with calculations carried out using the present model for smaller 
 clusters~\cite{rm_d_level}. 
Finally, we draw some conclusions.
\section{Microscopic tight-binding model}
\label{theo_model}

We consider a finite cluster of GaAs, where substitutional TM impurities 
 are introduced at Ga sites. The system is described 
 by a multi-orbital TB model, with parameters inferred from density functional 
theory (DFT) calculations~\cite{rm_d_level}. We include    
 $s$-, $p$- and $d$-orbitals for the impurity atoms while keeping 
 only $s$- and $p$-orbitals for the atoms of the host. This choice of the orbital basis 
 is motivated by DFT calculations, which show that the $d$-levels of Ga are located far below ($\approx$ 15~eV) the 
Fermi level~\cite{rm_d_level}. The Hamiltonian of the system is given by
\begin{equation}
H  =H_{\rm GaAs}+H_{\rm TM} + H_{\rm LRC}.%
\label{hamiltonian}
\end{equation}
The first term in 
Equation~(\ref{hamiltonian}) represents the TB Hamiltonian of the GaAs host, 
which can be further written as the sum of two terms
\begin{equation}
H_{\rm GaAs} =  H_{\rm band} + H_{\rm SOI}\;, 
\end{equation}
where
\begin{equation}
H_{\rm band}   =\sum_{ij,\mu\mu^{\prime},\sigma}t_{\mu\mu^{\prime}}^{ij}a_{i\mu\sigma
}^{\dag}a_{j\mu^{\prime}\sigma}\;,
\label{tb}
\end{equation}
is the $sp^3$ Slater-Koster Hamiltonian for bulk GaAs~\cite{chadi_prb77, slaterkoster_pr54,papac_jpcm03}, 
with parameters $t_{\mu\mu^{\prime}}^{ij}$ representing both on-site
energies and nearest-neighbors hopping  integrals. 
Here $a_{i\mu\sigma}^{\dag}$ and $a_{i\mu\sigma}$
are electron creation and annihilation operators; 
$i$ and $j$ are atomic indices that run over all atoms other than 
the impurity, $\mu$ and $\mu^{\prime}$
are orbital indices and $\sigma = \uparrow, \downarrow $ is a spin index defined
 with respect to an arbitrary quantization axis. 
The spin-orbit interaction (SOI) is introduced as an on-site one-body term
\begin{equation}
H_{\rm SOI}  =\sum_{i,\mu\mu^{\prime},\sigma\sigma^{\prime}}\lambda_{i}\langle\mu
,\sigma|\vec{L}\cdot\vec{S}|\mu^{\prime},\sigma^{\prime}\rangle a_{i\mu\sigma
}^{\dag}a_{i\mu^{\prime}\sigma^{\prime}}
\label{so}
\end{equation}
where $\lambda_i$ are the  
re-normalized spin-orbit splittings~\cite{chadi_prb77}.\\
The second term 
in Equation~(\ref{hamiltonian}) is the Hamiltonian of the TM impurity. 
We have
\begin{equation}
H_{\rm TM} =\sum_{i,m,\mu,\nu,\sigma}\big( t_{\mu\nu}^{im}a_{i\mu\sigma
}^{\dag}a_{m\nu\sigma} + 
t_{\mu\nu}^{im \star}a_{m\nu\sigma}^{\dag} a_{i\mu\sigma
}\big)
\nonumber\\
+\sum_{m,\nu,\sigma}\epsilon_{m \nu \sigma}
a_{m\nu\sigma}^{\dag}a_{m\nu\sigma}\nonumber\\
+ H^{\rm TM}_{\rm SOI}\;,
\label{imp}
\end{equation}
where $a_{m\nu\sigma}^{\dag}$ and $a_{m\nu\sigma}$ are creation and annihilation operators
at the impurity site $m$; the orbital index $\nu$ runs over $s$-, $p$-, and $d$-orbitals.
The first term in 
Equation~(\ref{imp}) describes the hopping between the impurity and its nearest-neighbors As atoms. 
For the TB hopping parameters between the impurity $d$-orbitals and the nearest-neighbor  
As $s$- and $p$-orbitals we use the same values as for the corresponding hopping parameters 
between Ga and As~\cite{Bassani_PRB_57_6493}.
The second term in Equation~(\ref{imp}) represents on-site energies of the impurity for a given orbital.
The $d$-orbital energies $\epsilon_{m\,d\,\sigma}$ play an important role in the model. 
Their values for ``spin-up'' (majority) 
 and ``spin-down'' (minority) electrons
are different, which leads to a different occupation for opposite spin states, 
and hence to a non-zero spin magnetic moment at the impurity site. 
As a first estimate of the on-site $d$-orbital energies, we use the values of the exchange-split 
majority and minority $d$-levels, which can be identified 
in the spin- and orbital-resolved density of states (DOS) of the impurity, calculated with DFT.  
For the exact parametrization of the TM impurity Hamiltonian the 
reader is referred to reference~\cite{rm_d_level}, where the model was first introduced. 
The last term in Equation~(\ref{imp}) is an on-site SOI term for the impurity atom,
analogous to the one given in  Equation~(\ref{so}). The SOI terms $ H_{\rm SOI}$ and $ H^{\rm TM}_{\rm SOI}$ 
will cause the total ground-state energy of the system 
to depend on the direction of the impurity magnetic moment, 
defined with respect to an arbitrary quantization axis. This is the origin of the magnetic anisotropy energy.\\
Finally, the third term 
in Equation~(\ref{hamiltonian}) 
\begin{equation}
H_{\rm LRC}  =\frac{e^{2}}{4\pi\varepsilon_{0}\varepsilon_{r}}\sum_{m}\sum_{i\mu\sigma
}\frac{a_{i\mu\sigma}^{\dag}a_{i\mu\sigma}}{|\vec{r}_{i}\mathbf{-}\vec{R}%
_{m}|}\;,%
\label{lrc}
\end{equation}
is a long-range repulsive Coulomb 
potential that is dielectrically screened by the host material
(the index $m$ runs over all impurity atoms), with $\varepsilon_r$ being the dielectric 
constant. This term prevents the charging of the impurity atom 
and localizes the acceptor hole around the impurity~\cite{scm_MnGaAs_paper1_prb09}.\\
The electronic structure of GaAs with a single substitutional Mn impurity atom 
is obtained by performing supercell-type calculations  
 with periodic boundary conditions applied in either 2 or 3 dimensions, depending on
whether we are studying the $\left(  110\right)  $ surface or a bulk-like
system. 
%The $\left(  110\right)  $ surface of GaAs is attractive from both
%theoretical and experimental points of view due to the absence of large surface 
%reconstruction. 
We do not take into account the 
modification in strain and relaxation caused by the presence
of the magnetic impurity. However
in order to remove 
artificial dangling-bond states that would otherwise appear in the
band gap, we include relaxation of surface layer positions, 
following a procedure put forward in 
Refs.~\cite{chadi_prl78} and \cite{chadi_prb79}.
For more details the reader is 
referred to reference~\cite{scm_MnGaAs_paper1_prb09}.\\
Based on our computational resources, we were able to  
fully diagonalize and obtain the entire eigenvalue spectrum of 
the Hamiltonian for clusters with up to 3200 atoms. 
For the clusters larger than 3200 atoms, 
we used the Lanczos method, built-in a commercial software 
package, MATLAB~\cite{MATLAB_2010}, which allowed us to compute eigenvalues 
in a narrow window of interest (typically few eigenvalues around the 
expected position of the Mn acceptor in the gap). The 
 outputs of the two methods were systematically compared 
to insure the stability 
 of the results against the variation of 
the diagonalization procedure.
\section{Results and discussion}
\label{results}
We present the results of calculations carried out for a single Mn dopant in bulk GaAs and near the 
(110) GaAs surface using the quantum-spin model, described in the previous section. The size of 
 the supercell in our TB model is varied between 3200 atoms, which is the maximum size that has 
 been investigated previously, to 30,000 atoms. 
  In general, our calculations produce the well-known features of Mn in the  bulk as well as on the 
surface, in agreement with previous theoretical work~\cite{scm_MnGaAs_paper1_prb09}.  
  However, as we show in the following, the model gives a better and more realistic estimate of 
  the Mn magnetic anisotropy energy and its dependence on the impurity position below the surface, as 
  the size of the cluster is increased.\\ 
\begin{figure}[htp]
\centering
\includegraphics[scale=0.245]{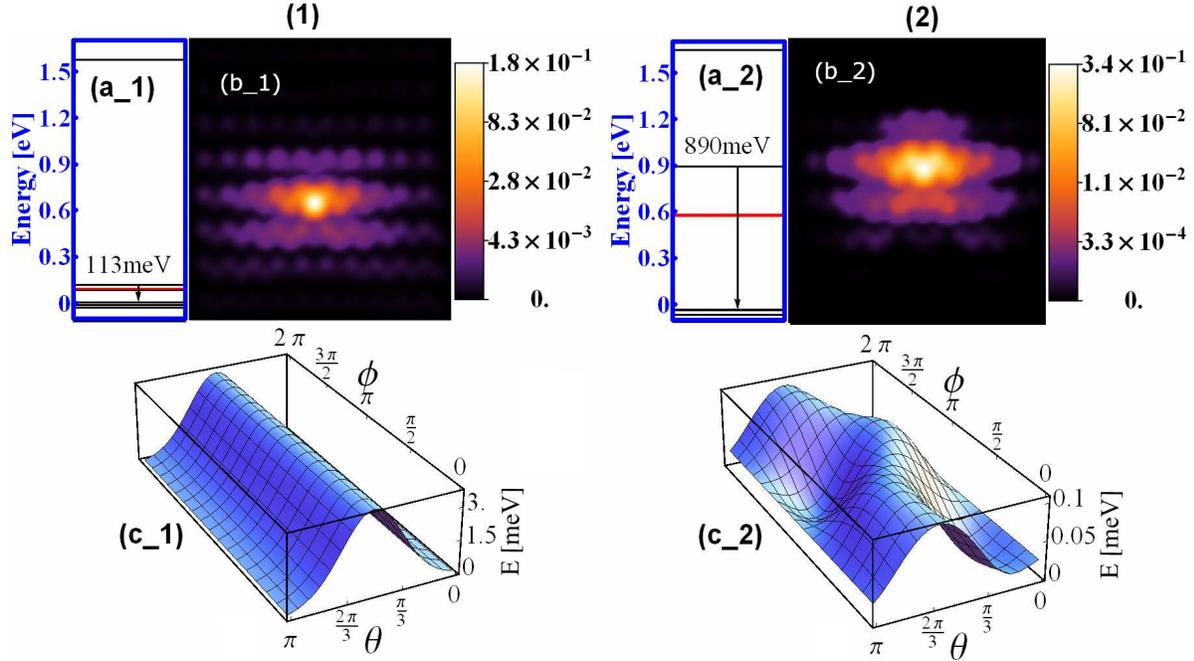}
\caption{Color online -- Electronic properties of a single Mn acceptor in GaAs, calculated using the TB model which 
 includes explicitly the Mn $d$-levels: panel (1) is for bulk and (2) for the (110) surface of GaAs. 
 (a\_1, a\_2) the eigenvalue spectrum, (b\_1, b\_2) the calculated LDOS for the acceptor state
 and (c\_1, c\_2) the magnetic anisotropy landscape of Mn. 
In panels (a\_1) and (a\_2) the red lines mark the highest occupied level while 
the black lines mark the acceptor state (the first level above the highest 
occupied level). Here we use the coordinate system with $\theta=0$ parallel
to the [001] direction and ($\theta=\frac{\pi}{2}$, $\phi=\frac{\pi}{2}$) parallel to [010] direction.}
\label{fig:spec-LDOS}
\end{figure}
Figure~\ref{fig:spec-LDOS} shows the in-gap electronic structure, 
the acceptor LDOS and the anisotropy landscape for Mn in the bulk 
(left panels) and on the (110) surface 
(right panels) of GaAs. 
Angles $\theta$ and $\phi$ used in panels (c\_1) and (c\_2) and throughout 
the paper, are defined such way that $\theta=0$ is parallel
to the [001] and ($\theta=\frac{\pi}{2}$, $\phi=\frac{\pi}{2}$) is parallel to [010] crystalline directions.
As it is shown in panels (a\_1) and (a\_2), Mn introduces three levels in the GaAs gap, 
with the highest level, which is unoccupied, 
known as the hole-acceptor level. The other two levels are occupied and they lie below the acceptor. 
The position of the acceptor level with respect to the valence band is found 
at 113~meV for the bulk, which reproduces exactly the experimental 
value~\cite{schairer_prb74, lee_ssc64, chapman_prl67, linnarsson_prb97}, and at 0.89~eV for the surface dopant, which 
 is also close to the experimental result~\cite{yazdani_nat06}. 
As one can see from Figure~\ref{fig:spec-LDOS}(b\_2), the calculated LDOS for the Mn acceptor on the surface   
shows more concentration of the spectral weight on the impurity site compared to the bulk case, which signals    
 a deeper and a more localized character of the acceptor state on the surface.\\
%In general, the calculations presented in Figure~\ref{fig:spec-LDOS} support the results 
%of the classical spin model~\cite{tangflatte_prl04, scm_MnGaAs_paper1_prb09} 
%and are in good agreement with other theoretical and 
%experimental results \cite{Jancu_PRL_08, yazdani_nat06}.
We would like to comment on one important feature of the calculated  
electronic structure of Mn acceptor in bulk GaAs.  
According to the calculation for a typical 3200-atom supercell [see Figure~\ref{fig:spec-LDOS}(a\_1)],  
the three levels introduced by Mn in the bulk GaAs gap 
are found to be spread over an energy interval of approximately $30$~meV, 
when SOI is included in the calculation [note that in Figure~\ref{fig:spec-LDOS}(a\_1), 
the top-most and the lowest levels in the gap are split by $\approx 30$~meV].\\
Figure \ref{fig:beyond_so} shows similar calculations for different supercell sizes. 
As one can see from the figure, the position of the three levels in the gap starts to shift  
as the supercell size is increased, gradually approaching saturation as a function of the size 
(the absolute position of the 
 levels does not change appreciably for clusters containing more than 20,000 atoms).
 However, the splitting 
between the levels as well as the relative position of the acceptor level with respect to the top of valance 
band remains unchanged (113~meV).
This is a shortcoming that the present quantum-spin model shares with 
the classical-spin models introduced in \cite{tangflatte_prl04} and \cite{scm_MnGaAs_paper1_prb09}. 
In fact the three levels of predominantly   
$p$-character, appearing in the gap, should be degenerate 
in the perfectly tetragonal environment of an impurity in bulk GaAs~\cite{JPCM_206005}. The lifting of the 
degeneracy is most likely related to the breaking of rotational symmetry 
 due to the essentially mean-field nature of the approximation for the exchange coupling between
the TM impurity $d$-levels and the $p$-levels of the nearest neighbor As atoms,  
 used in both the quantum- and the classical-spin model. Note that the same problem occurs in the DFT calculations, 
 which are also based on a broken-symmetry approach.  
In contrast to this, a perfectly three-fold degenerate level is expected for the
present model as well as for the classical spin model~\cite{tangflatte_prl04, scm_MnGaAs_paper1_prb09}, 
when SOI is switched off. We confirm this by calculating the in-gap level structure for Mn in the bulk in the absence of 
 SOI. We find that the splitting between the levels reduces from 11.54~meV 
for a 3200-atom to 0.62~meV for a 20,000-atom cluster (Figure~\ref{fig:Beyond}). That is, 
in the absence of SOI the splitting between the three Mn-induced levels in the bulk GaAs gap 
is zero for this model.\\
\begin{figure}[htp]
\centering
\includegraphics[scale=0.18]{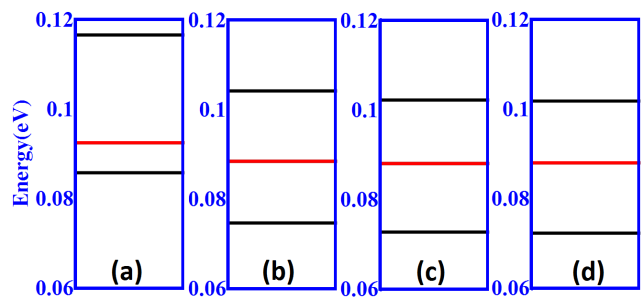}
\caption{Color online -- The in-gap electronic level structure for a single Mn acceptor in bulk GaAs when SOI is included, 
 calculated for different size of the supercell: (a) 3200 atoms, (b) 9900 atoms, (c) 20216 atoms and (d) 30976 atoms. The red line indicates the position of the Fermi level.}
\label{fig:beyond_so}
\end{figure}
\begin{figure}[htp]
\centering
\includegraphics[scale=0.2]{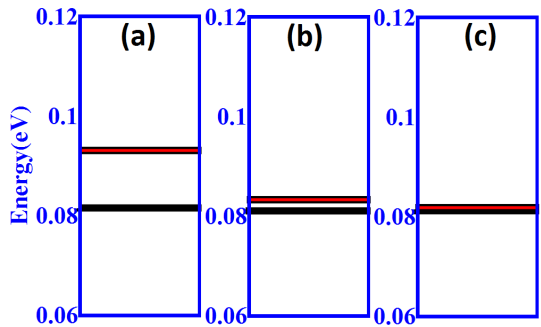}
\caption{Color online -- The in-gap electronic level structure for a single Mn acceptor in bulk GaAs when SOI is \textit{not} included, 
 calculated for different size of the supercell: (a) 3200 atoms, (b) 9900 atoms, and (c) 20216 atoms. The red line indicates the position of the Fermi level.}
\label{fig:Beyond}
\end{figure}\\
To summarize, our calculations show that 
(\textit{i}) in the presence of SOI the splitting between the three levels in the gap, as well as the 
relative position of the acceptor level with respect to the top 
of the valence band (113~meV) remain unchanged even for very large clusters 
 containing up to 30,000 atoms (Figure~\ref{fig:beyond_so}), and   
(\textit{ii}) in the absence of SOI the small splitting between the levels, 
which is still present in calculations for a 3200-atom supercell, is 
purely due to a finite-size effect and  
 quickly vanishes with increasing the 
size of the supercell (Figure~\ref{fig:Beyond}).\\
We will now focus on the calculations of magnetic anisotropy energy for 
 a single Mn in the bulk and on the (110) surface of GaAs. 
 In particular, we will discuss the trend of magnetic anisotropy with increasing 
 the size of the supercell.\\
In order to evaluate the magnetic anisotropy of the system, one should in principle
calculate the entire eigen-spectrum of the Hamiltonian. 
However, for larger clusters we are forced to use the Lanczos diagonalization method 
that allows us to obtain eigenvalues (and eigenvectors) of the Hamiltonian only 
in a very small window around the Fermi level (or around the position of the acceptor level in the gap). 
In the case  of the classical-spin model~\cite{scm_MnGaAs_paper1_prb09} one can overcome 
this difficulty by using the following important property of the system. 
It can be  shown that the 
energy of the (single-particle) acceptor level 
$\epsilon_{\rm acc}(\theta, \phi)$ and the (many-particle) ground state (GS) 
energy of the system $E(\theta, \phi)$ are very accurately related by the following expression 
\begin{equation}
\epsilon_{\rm acc}(\theta, \phi) 
= - E(\theta, \phi) + C\,,
\label{acc_anisotropy}
\end{equation}
where $C$ is a constant independent of $\theta$ and $ \phi$.
This means that the sum of the two energies $E(\theta, \phi)$ and $\epsilon_{\rm acc}(\theta, \phi)$ is the same for  
any direction of the Mn magnetic moment. If $(\theta_{\rm max},  \phi_{\rm max})$  
and $(\theta_{\rm min},  \phi_{\rm min})$ define the two directions
where $E(\theta, \phi)$ attains its maximum and minimum value respectively,
from Equation~\ref{acc_anisotropy} we obtain
\begin{equation}
%{\rm MAE}\equiv 
[E(\theta_{\rm max}, \phi_{\rm max})- E(\theta_{\rm min}, \phi_{\rm min})]  \nonumber\\
+  [\epsilon_{\rm acc}(\theta_{\rm max}, \phi_{\rm max})- \epsilon_{\rm acc}(\theta_{\rm min}, \phi_{\rm min})] =0\;.
\label{Delta_mae}
\end{equation}
The quantity $[E(\theta_{\rm max}, \phi_{\rm max})- E(\theta_{\rm min}, \phi_{\rm min})] $
is by definition the magnetic anisotropy energy (${\rm MAE} $) of the system. Similarly,  Equation~\ref{acc_anisotropy} implies that
$[\epsilon_{\rm acc}(\theta_{\rm max}, \phi_{\rm max})- \epsilon_{\rm acc}(\theta_{\rm min}, \phi_{\rm min})]$
is the {\it opposite} of the magnetic anisotropy of the acceptor level, $(-{\rm MAE})_{\rm acc}$.
Therefore, we can rewrite Equation~\ref{Delta_mae} as 
\begin{equation}
\Delta_{\rm  MAE} \equiv {\rm MAE} - ({\rm MAE})_{\rm acc}= 0
\label{Delta_mae2}
\end{equation}
Equations~\ref{acc_anisotropy} and \ref{Delta_mae2} contain a very strong physical result and are particularly 
 useful for practical calculations of the magnetic anisotropy energy for large clusters, namely   
they imply that the total anisotropy 
of the system is essentially determined by the anisotropy of the single-particle acceptor level. 
This picture remains valid as long as the coupling to the conduction band is not sensitive 
to the magnetization orientation.\\
In contrast to the classical-spin model, the results of the calculations of magnetic anisotropy energy 
 using the quantum-spin model indicate that  
Equation~\ref{acc_anisotropy} is, in principle, not satisfied. 
As a result, the quantity $\Delta_{\rm  MAE}$ is not exactly zero in our calculations, however 
 its value is negligibly small. 
 We suggest that this small change in the difference between the GS and the acceptor
anisotropy energies is due to the inclusion of the $d$-orbitals, which 
brings about a magnetization-direction dependent coupling with the conduction band.
In the classical-spin model, the polarized spins of the majority 
$d$-electrons are represented by a classical vector with a fixed 
magnitude of $+5/2$ $\mu _B$. This only affects the (occupied) 
energy-levels of the valence band through its SOI-induced orientation 
dependence. In contrast, our quantum-spin model includes the impurity 
$d$-orbitals and the corresponding hopping amplitudes between the 
$d$-orbitals and the nearest neighbor As atoms explicitly in the 
Hamiltonian. Unoccupied minority $d$-levels, located way up in the 
conduction band, hybridize with like-spin As $p$-orbitals of the 
valence band. This coupling is responsible for the small deviation 
from Equation~\ref{Delta_mae2}, which is also affected by the distance of the Mn atom from 
the surface. The fact that the deviation from the classical-spin model 
result is so small indicates that the effect of the conduction 
band hopping in this system is not very important.\\
In Figure~\ref{fig:MAE_size} we present the calculated magnetic anisotropy energy for the Mn acceptor in the bulk, 
for very large clusters containing up to approximately 34,000 atoms. 
These calculations show explicitly that the bulk magnetic anisotropy energy decreases 
 with increasing the cluster size, dropping drops from 3.7~meV for a 3200-atom cluster to
 the very small value of 0.09~meV for a cluster containing 34,000 atoms.
 The inset in Figure~\ref{fig:MAE_size} shows that the magnetic anisotropy energy decreases linearly with the inverse
 number of atoms in the cluster.
\begin{figure}[htp]
\centering
\includegraphics[scale=0.5]{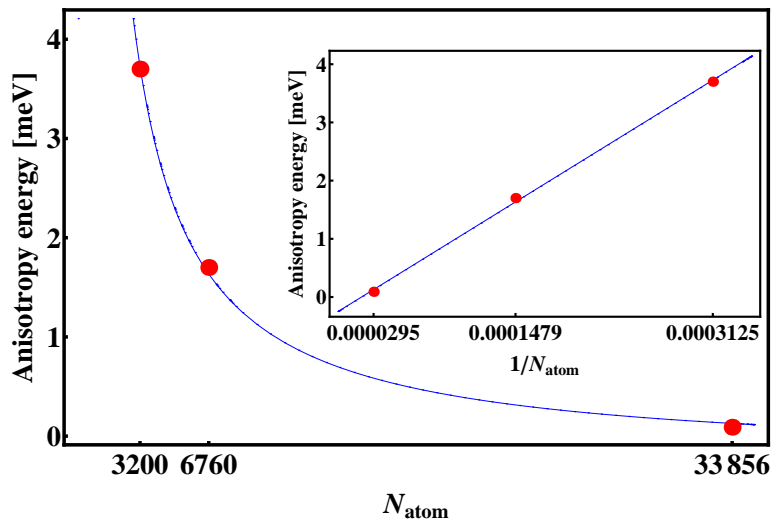}
\caption{Color online -- The acceptor magnetic anisotropy energy 
(MAE$_{\mathrm{acc}}$) of Mn impurity in the bulk 
for three different sizes of the supercell   
as a function of number of atoms in the cluster ($N_{atom}$). 
The inset shows the magnetic anisotropy energy as a function of the inverse number of atoms.
The solid line is obtained by fitting to the discrete data points.}
\label{fig:MAE_size}
\end{figure}\\
Figure~\ref{fig:MAE-Layers} shows the calculated magnetic anisotropy energy of the system as 
a function of the Mn depth, for two different cases: (\textit{i}) a 3200-atom supercell with   
 19 Ga layers along the [110] direction, and (\textit{ii}) a 6760-atom supercell with   
 25 Ga layers. In case (\textit{i}), when the system size is still suitable for 
 full diagonalization, we performed systematic comparison between exact calculation 
 of MAE, based on the entire eigen-spectrum of the Hamiltonian, and the Lanczos result, 
 obtained by calculating the acceptor anisotropy, MAE$_\mathrm{acc}$. We find that the two sets 
 of calculations, in particular the value of the magnetic anisotropy energy for bulk, surface and subsurfaces, 
 are in good agreement with each other and with the results of the classical-spin model~\cite{scm_MnGaAs_paper1_prb09, rm_d_level}. 
 This suggest that the deviations from Equation~\ref{Delta_mae2} are indeed small even when the $d$-levels of the impurity are 
  included explicitly in the Hamiltonian. 
 The only discrepancy is found in the magnetic anisotropy landscapes, calculated for the surface and the first sublayer, which 
  will be discussed later in the text. In case (\textit{ii}) the full diagonalization results are no longer available and 
  we rely solely on the calculations of MAE$_\mathrm{acc}$. Note that for a 3200-atom cluster, the 9-th sublayer 
  corresponds to the center of the cluster and the Mn atom can not be positioned any further away from the surface. 
  However, for a 6760-atom cluster, we are able to perform calculations for Mn in sublayers 1 to 12 below the surface. 
  This enables us to draw more general conclusions on the trend of magnetic anisotropy for Mn positioned in the sublayers. \\
\begin{figure}[htp]
\centering
\includegraphics[scale=0.5]{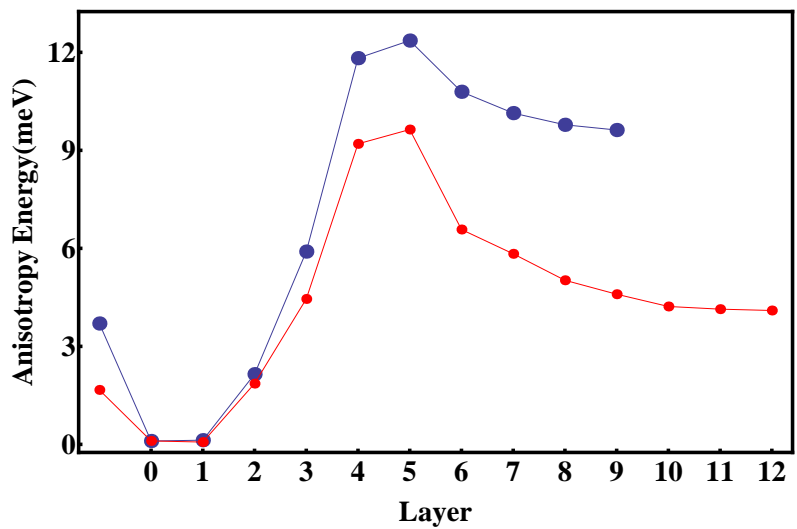}
\caption{Color online -- Magnetic anisotropy energy as 
a function of the Mn depth. The values on the horizontal axis correspond to the sublayer index in which the Mn impurity 
 is located. The value at zero is the magnetic anisotropy energy of the (110) surface. The value   
 before zero of the horizontal axis shows the magnetic anisotropy energy of the bulk. 
Blue curve is for the case of 3200 atoms (19 Ga layers) and red curve 
is for the case of 6760 atoms (25 Ga layers) in the cluster.}
\label{fig:MAE-Layers}
\end{figure}
We will now discuss some of the key features of the magnetic anisotropy calculations for Mn positioned in the sublayers 
(Figure~\ref{fig:MAE-Layers}). For both cluster sizes considered here, 
the anisotropy energy increases as Mn is moved down to the fifth sublayer  
and it decreases for Mn positioned  deeper below the surface. This peculiar behavior 
has been reported previously in calculations based on the classical-spin model~\cite{scm_MnGaAs_paper1_prb09, rm_d_level}. 
 It can be explained based on the following arguments. 
%these statements have been removed in the revised version
%As the impurity is moved away from the first sublayer, the wavefunction of the corresponding acceptor 
% state becomes more extended~\cite{scm_MnGaAs_paper1_prb09, rm_d_level} and will be therefore 
% strongly affected by the surface, until the Mn atom is moved 
%deep enough so that the surface effects become negligible (this situation corresponds to the sixth sublayer). 
%A very small magnetic anisotropy ($\approx$0.06 meV) of the first sublayer is due to the highly localized 
% and less anisotropic character of the acceptor wavefunction, compared to 
%the acceptor in the surface layer~\cite{rm_d_level}.
A very small magnetic anisotropy of the surface layer ($\approx$0.11 meV) and first 
sublayer ($\approx$0.06 meV) is due to the highly localized 
and less anisotropic character of the acceptor wavefunction, compared to 
the acceptor in other sublayers~\cite{rm_d_level} [Panels (a) and (b) of Figure~\ref{fig:LDOS}]. 
As the impurity is moved away from the first sublayer, the wavefunction of the corresponding acceptor 
 state becomes more extended~\cite{scm_MnGaAs_paper1_prb09, rm_d_level} and will be therefore 
 strongly affected by the surface, until the Mn atom is moved 
deep enough so that the surface effects become negligible (this situation corresponds to the sixth sublayer).
Furthermore, we point out another important feature of Figure~\ref{fig:MAE-Layers}, 
which has not been discussed previously and 
partly motivates the calculations for larger clusters. As the Mn impurity is moved down towards the center of the cluster,  
one would expect the anisotropy energy to decrease until it reaches its bulk value,  
 when Mn is placed in the center of the cluster. Based on the calculation for a relatively small 3200-atom supercell   
(blue curve in Figure~\ref{fig:MAE-Layers}), it is not clear whether this is indeed the case. In this calculation, not only the bulk 
 anisotropy energy is non negligible (3.7~meV) but also the anisotropy for Mn in the 9th sublayer is quite large ($>$9~meV). 
 This issue is clarified by the calculation for a larger cluster containing 6760 atoms 
(red curve in Figure~\ref{fig:MAE-Layers}). Firstly, the 
 maximum value of the magnetic anisotropy energy, which occurs for the Mn in the fifth sublayer, decreases compared to the smaller cluster, which 
 is also consistent with the bulk calculations (Figure~\ref{fig:MAE_size}). Secondly, 
 the anisotropy energy decreases even further as Mn is moved  away from the surface  
 beyond the 9-th sublayer. These observations confirm the trend towards saturation of the 
   magnetic anisotropy energy to its bulk value, as the impurity is positioned in deeper sublayers.\\
We would like to add a remark about the trend of the anisotropy 
energy as a function of cluster size. When Mn is placed in the 
middle of the cluster in the bulk calculation, the distance between 
the two Mn atoms in the neighboring supercells and, as a result, 
the overlap between their wavefunctions depend on the size of 
the supercell. With increasing the size of the cluster, the Mn 
wavefunction will eventually approach that of an isolated impurity 
and, as a result of perfect tetragonal symmetry, its MAE drops to a 
value very close to zero (exactly zero in the limit of an infinitely 
large cluster). In the case of Mn placed in the sublayers, as long 
as it is not very far from the surface (for example, in the fifth 
sublayer), the situation is different from the bulk. Increasing 
the size of the supercell does finally detach the Mn wavefunction 
from the boundary planes perpendicular to (110) surface. However, 
the Mn-acceptor wavefunction along the [110] direction will still 
be influenced by the lower symmetry of (110) surface. No matter 
how large the supercell is, the MAE of Mn in the sublayers  (say, 
in the fifth sublayer) will not vanish as long as this sublayer 
remains close enough to the  surface. Indeed, our new set of calculations 
for a cluster containing more than 55,000 atoms with 54 Ga layers 
along the [110] direction confirmed our prediction. Due to 
computationally-demanding and time-consuming nature of these 
calculations, we were not able to rotate the quantization axis for 
all possible directions to plot a figure like 
Figure~\ref{fig:MAE-Lay_beyond}. Instead, we chose two 
specific directions, known to be the easy and the hard directions, 
for sublayers number 5 and 26. Note that, the 26 sublayer is the 
deepest sublayer for this cluster. Therefore, its magnetic anisotropy 
landscape should resemble the 12th sublayer of a 6760-atoms cluster, 
shown in Figure~\ref{fig:MAE-Lay_beyond}, or the 9th sublayer of a 3200-atoms 
cluster, shown in Figure 9 of the reference~\cite{scm_MnGaAs_paper1_prb09}. (We assume that the 
anisotropy landscape will not change qualitatively with increasing the 
cluster size, therefore the easy and hard directions will be the same). 
In these new calculations, we find that the anisotropy energy between 
the easy and hard directions for the 5th sublayer is slightly smaller 
(9.1 meV) than value found for a smaller cluster (9.6 meV from Figure~\ref{fig:MAE-Layers}). 
However, for the 26th sublayer, the anisotropy energy is only 0.05 meV 
(compare to 4.1 meV for the sublayer in the middle of 6760-atoms cluster (12th)). 
These calculations further confirm that the wavefunction of the Mn atom 
placed in the middle of a very large cluster will be highly isolated from 
all boundaries and its anisotropy energy will drop to zero, while for the 
sublayers close to the surface it will remain anisotropic.\\
Figure~\ref{fig:MAE-Lay_beyond} shows the acceptor magnetic anisotropy landscape
 for Mn near the (110) GaAs surface, calculated for a 6760-atom cluster. Here 
 the magnetic anisotropy energy is plotted for different directions of the 
  Mn spin quantization axis, characterized by angles $\theta$ and $\phi$. According to  
  previous calculations~\cite{rm_d_level}, for the directions considered here ($\theta\in[0,\pi]$, $\phi\in[0,\pi]$), 
 the impurity magnetic moment has one easy and one hard axis.\\
\begin{figure}[htp]
\centering
\includegraphics[scale=0.38]{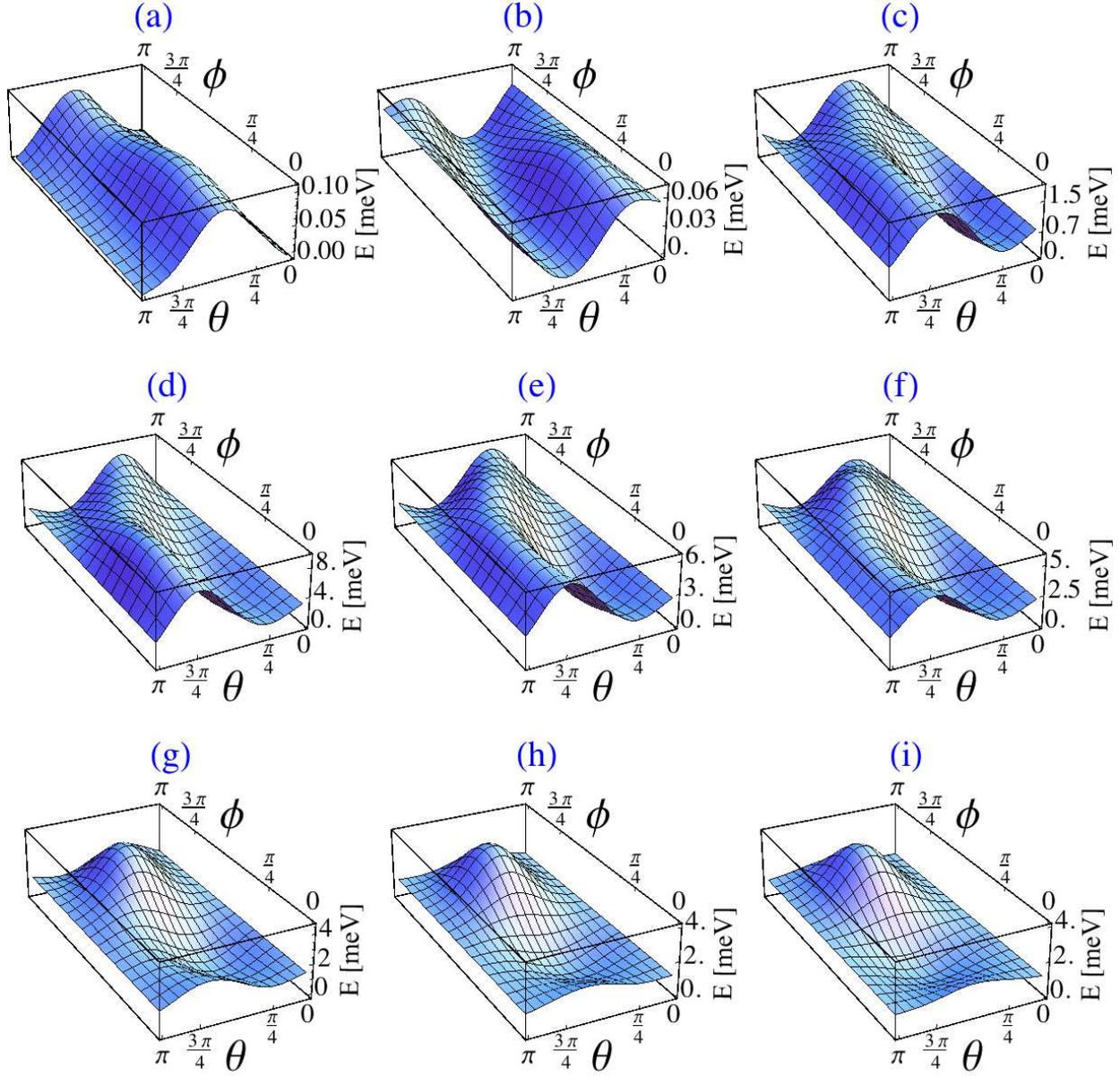}
\caption{Color online -- Magnetic anisotropy energy landscape as 
a function of the Mn depth for the supercell containing 6760 atoms. Panels 
(a)-(i) correspond to the Mn atom being placed on the surface and in sublayer number 
1, 2, 5, 6, 7, 9, 11 and 12, respectively. Here we use the coordinate system with $\theta=0$ parallel
to the [001] direction and ($\theta=\frac{\pi}{2}$, $\phi=\frac{\pi}{2}$) parallel to [010] direction.}
\label{fig:MAE-Lay_beyond}
\end{figure}
The overall pattern 
of the magnetic anisotropy landscape for this cluster size closely resembles 
the previous calculations for smaller clusters~\cite{rm_d_level}, with only two exceptions. 
The anisotropy landscapes (but not the absolute value of the magnetic anisotropy energy) 
 for Mn on the surface and in the first sublayer [Figure~\ref{fig:MAE-Lay_beyond}(a) and (b)] 
is different from those reported in \cite{rm_d_level}. 
As explained earlier, in the present model, which includes explicitly the 
$d$-levels of the impurity atom, the magnetic anisotropy energy of the system 
  is not necessarily equal (in absolute value) to the 
 anisotropy of the single-particle 
acceptor level. In particular, if the anisotropy energy itself is small, which 
  is indeed the case for the surface and the first sublayer, the difference 
  between MAE and MAE$_\mathrm{acc}$ can become visible for different direction of 
  the impurity magnetic moment. In fact, we carefully compared 
  the acceptor and the GS anisotropy landscapes in all sublayers 
  for the smaller cluster size. We find that the difference is indeed most visible 
   for Mn on the surface and in the first sublayer, which further supports our 
   calculations for the larger cluster, where calculations of the GS anisotropy are 
    not possible.\\
\begin{figure}[htp]
\centering
\includegraphics[scale=0.32]{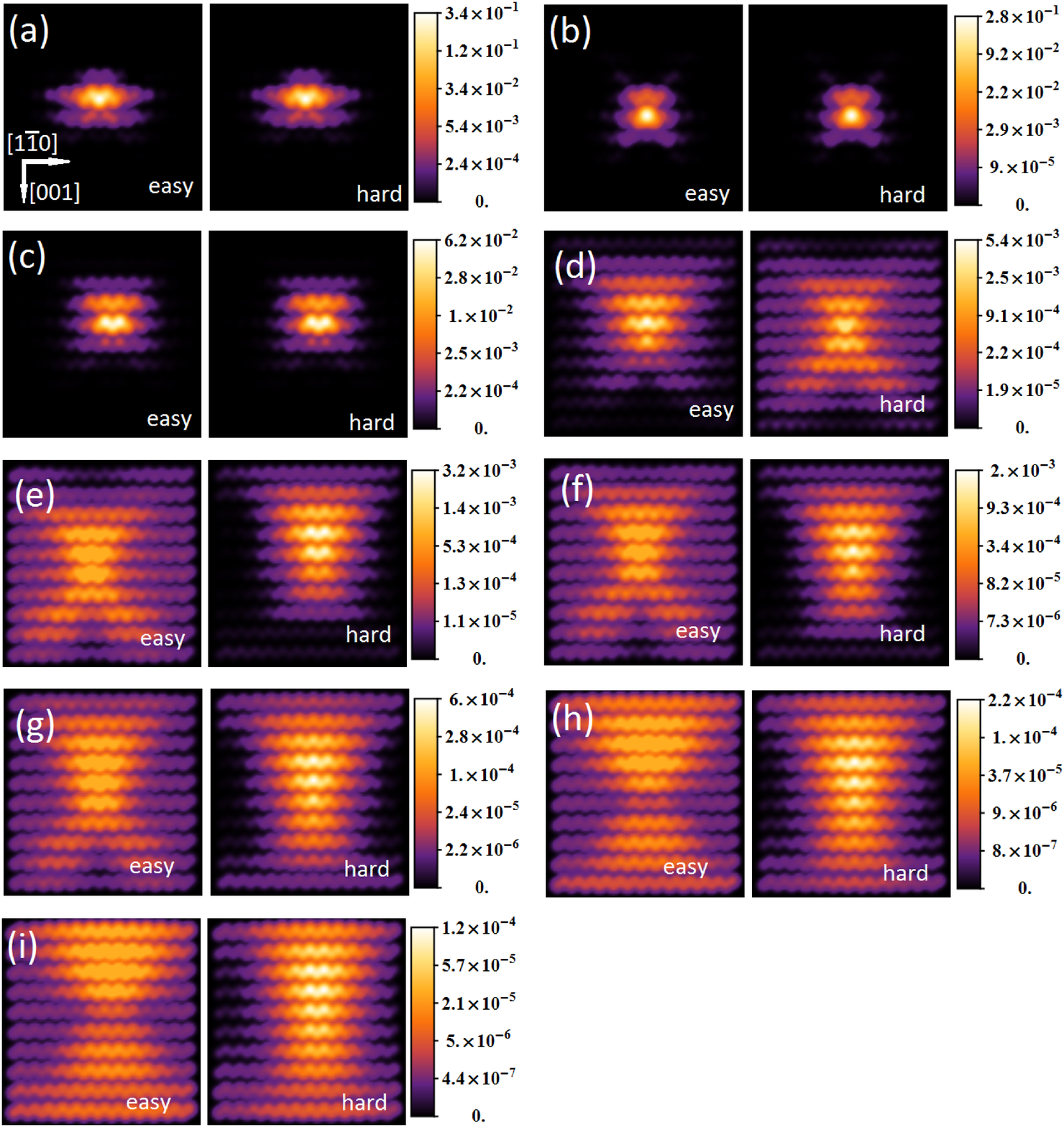}
\caption{Color online -- (110) Mn-acceptor LDOS as 
a function of the Mn depth for the supercell containing 6760 atoms. Panels 
(a)-(i) correspond to different Mn depth, similar to Figure~\ref{fig:MAE-Lay_beyond}. In this figure, \textit{easy} or \textit{hard}
refer to the direction of the quantization axis for which the GS energy is minimum or maximum, respectively.}
\label{fig:LDOS}
\end{figure}
In order to be able to address STM topographies of the Mn acceptor,
when Mn is placed on the surface or sublayers of GaAs (110) surface, 
we plotted the corresponding LDOS in Figure~\ref{fig:LDOS}.
In agreement with previous theoretical and experimental 
works~\cite{Jancu_PRL_08, scm_MnGaAs_paper1_prb09, PhysRevB.77.075328, rm_d_level},
we observed the following properties in the LDOS.
All images show a $(1\bar{1}0)$ mirror plane as reported previously.
The LDOS extends spatially even further along the [001] direction,
for the case in which Mn is placed in deeper sublayers~\cite{PhysRevB.77.075328, scm_MnGaAs_paper1_prb09}.
The deeper the Mn is from the (110) surface, the more symmetric the anisotropic features
introduced by Mn becomes, which is an indication that the environment for this Mn depth is more 
resembling the bulk host material. 
Using larger clusters, we were able to position the impurity as deep as in the 12th sublayer. As one can see from the last two panels of Figure 8, the symmetric butterfly shape becomes more pronounced in 11th and 12th sublayers~\cite{PhysRevB.77.075328, scm_MnGaAs_paper1_prb09}.
The acceptor-LDOS in deeper sublayers show a triangular shape, which shifts to a butterfly shape
with stronger upper-wing as Mn moves deeper. This has been described to be related 
to the intrinsic strain associated with the buckling relaxation~\cite{Jancu_PRL_08}.
The increase in the intensity 
along the hard axis compared to the easy axis after the fifth sublayer
has been observed and explained previously~\cite{scm_MnGaAs_paper1_prb09}.
The more pronounced LDOS for the deeper 
sublayers compared to previous results is due to 
the smaller magnetic anisotropy barrier between the easy and hard axes.
The deep acceptor for Mn on the surface and in the first sublayer results in highly localized LDOS, as seen in Figure 8 (a) and (b), respectively. As the Mn atom is moved down towards the center of the cluster, its binding energy decreases and the associated wavefunction is less localized [see Figure 8 (c)-(i)]. Note that, although  the shape of the acceptor LDOS in the second sublayer [Figure 8 (c)] appears to be less extended compared to subsequent layers, only 10\% of the spectral weight is located on the Mn atom, indicating a more delocalized acceptor wavefunction compared to the surface and the first sublayer. 
Finally, due to the larger size of the cluster, one can see that the wavefunction for localized cases
is completely detached from the boundary planes perpendicular to the (110) plane.
\section{Conclusions}
In this paper we have investigated the in-gap electronic structure 
and the magnetic anisotropy energy of a 
 single Mn acceptor in bulk and near the (110) surface of GaAs, 
 using a fully-microscopic TB model with supercells  
   containing up to $3.4\cdot 10^4$ atoms. 
The main outstanding issue addressed in our calculations has been 
the effect of the finite supercell size on the 
degeneracy of the impurity-induced energy levels in the bulk GaAs with and 
without SOI, and on the behavior of the magnetic anisotropy 
energy as a function of the Mn depth from the surface.\\
We found that in the absence of SOI, the three acceptor energy levels, 
introduced by the Mn dopant in the bulk GaAs gap, become degenerate with  
increasing the cluster size, which is expected from symmetry arguments. 
However, in the presence of SOI, the finite splitting between the levels, which is  
of the order of 30~meV, remains unchanged with increasing the cluster size 
up to $3.4\cdot 10^4$ atoms. We attribute this effect to the 
shortcomings of the mean-field treatment of the exchange coupling between 
the Mn impurity spin and its nearest neighbor As atoms.\\ 
The calculations of the magnetic anisotropy energy for Mn in bulk and near the surface 
revealed a number of important features, which have not been investigated previously. 
In particular, we showed for the first time that the non-negligible anisotropy of the 
Mn dopant in the bulk, found in earlier calculations, is due to a finite-size effect 
and that it indeed vanishes with increasing the size of the supercell. 
We also found that the magnetic anisotropy for Mn near the surface decreases 
considerably for larger clusters. A clear tendency of the surface anisotropy towards 
the bulk value was observed, as the dopant was moved away from the surface. 
In addition, based on the calculations of magnetic anisotropy,  
we identified some important differences between the present treatment, which takes 
into account the impurity $d$-levels, and the classical-spin model, which 
treats them as an effective classical spin. It was shown that, in the former case 
the robust relationship between the ground states anisotropy  and  
the acceptor anisotropy no longer holds, due to the explicit inclusion of the impurity 
$d$-levels in the Hamiltonian. 
In conclusion, our calculations provide an accurate and detailed picture of the electronic structure, LDOS  
and magnetic anisotropy for a single Mn dopant, positioned in the vicinity of the (110) GaAs surface.  
We anticipate that these result will be important for interpreting the on-going STM experiments on this and other 
similar TM-impurity systems, and in particular for on-going experimental efforts to 
manipulate the Mn acceptor states by means of external electric and magnetic fields.
A reliable estimate of the magnetic anisotropy landscape for individual TM  dopants close to the surface, like the one 
presented here, is also essential to extract an effective spin Hamiltonian for the impurity spin, following
for example the procedure put forward in reference \cite{scm_MnGaAs_paper3_prl011}.
Effective spin Hamiltonians for solitary TM dopants in a semiconductor host can be used to model magnetic excitations,
which are probed in spin inelastic electron tunneling spectroscopy\cite{Khajetoorians2010}.
%QUI
%
%
\ack 
This work was
supported by the Faculty of Technology
at Linnaeus University, by the
Swedish Research Council under Grant Number: 621-2010-3761, 
and the NordForsk research network 080134 
``Nanospintronics: theory and simulations".
Computational resources have
been provided by the Lunarc center for scientific and 
technical computing at Lund University.
%
%
% The \nocite command causes all entries in a bibliography to be printed out
% whether or not they are actually referenced in the text. This is appropriate
% for the sample file to show the different styles of references, but authors
% most likely will not want to use it.
%\nocite{*}
\section*{References}
\bibliographystyle{iopart-num}
\bibliography{larg_clus}% Produces the bibliography via BibTeX.
%
%\bibliography{refs}% Produces the bibliography via BibTeX.
%
%\section*{References}
%\begin{thebibliography}{99} 
%
%\bibitem{Hasan} Hasan M Z and Kane C L 2010 \textit{Rev. Mod. Phys.} \textbf{82} 3045  
%\bibitem{XLQi} Qi X-L and Zhang S-C 2011 \textit{Rev. Mod. Phys.} \textbf{83} 1057   
%
%\end{thebibliography}
%
\end{document}